# Large Mode Volume Integrated Brillouin Lasers for Scalable Ultra-Low Linewidth and High Power

Kaikai Liu[1], Karl D. Nelson[2], Ryan O. Behunin[3, 4], Daniel J. Blumenthal[1*]

[1]Department of Electrical and Computer Engineering, University of California Santa Barbara, Santa Barbara, CA, USA
[2]Honeywell Aerospace, Plymouth, MN, USA
[3]Department of Applied Physics and Materials Science, Northern Arizona University, Flagstaff, Arizona, USA
[4]Center for Materials Interfaces in Research and Applications (¡MIRA!), Northern Arizona University, Flagstaff, AZ, USA
[*]Corresponding author (danb@ucsb.edu)

## Abstract

Generating ultra-low linewidths and high output power in an integrated single mode laser remains a critical challenge for future compact, portable, precision applications. Moreso, achieving these characteristics in a laser design that enables scaling to lower linewidths and higher power, and implementation in a wafer-scale integration platform that can operate from the visible to near-IR and be integrated with other components. Such an advance could impact a wide array of applications including atomic and quantum sensing and computing, metrology, coherent fiber communications and sensing, and ultra-low-noise mmWave and RF generation.  Yet, achieving these goals in an integrated laser has remained elusive.  Here, we report a class of integrated laser that overcomes these limitations, with demonstration of a 31 mHz instantaneous linewidth, the lowest linewidth to date to the best of our knowledge, with 41 mW output power and 73 dB sidemode suppression ratio and can be tuned across a 22.5 nm range. This performance is possible due to Brillouin nonlinear dynamics that occurs within an large mode volume, nonlinear photon-phonon, MHz-scale-FSR, ultra-low loss silicon nitride resonator cavity.  This laser design can scale to a new operating regime of mHz fundamental linewidth and Watt class lasers. Such lasers hold promise to unlock new sensitivity and fidelity for precision quantum experiments, portable precision applications, and atomic, molecular, and optical physics.

# Introduction

Low fundamental linewidth and low-noise integrated lasers with high output power are critical for a wide range of precision scientific and commercial applications, including coherent communications[1], fiber-optic sensing[2], coherent laser ranging[3], atomic and quantum sensing[4], atomic clocks[5], and mmWave generation[6,7]. Specific examples include delivering precision high power light to narrow atomic transitions, for example preparation of cesium and rubidium Rydberg states for neutral atom computing[45,46] generation of extremely low phase noise mmWave and RF sources[6], and precision neutral atom gravitational time-shift measurements[47]. Enabling novel laser designs that enable the scaling to continually reduced linewidth and increased output power will unlock the potential for new precision applications and miniaturization of precision lab-scale lasers will improve experimental reliability and portability. Yet today, integrated laser solutions that can deliver this class of performance have remained elusive. There is a need for integrated lasers that can achieve the operating regime of low fundamental linewidth and high output power and can provide scaling towards mHz fundamental linewidths and Watt class output power.

Lab-scale large mode volume lasers, such as fiber-based external cavity lasers (ECLs), deliver high performance via a large intra-cavity photon number and high cavity quality factor (Q) needed to reduce linewidth and increase output power. Yet these table-top lasers are highly susceptible to environmental disturbances and require complex cavity stabilization techniques, with linewidths largely limited to the Hz level regime. Photonic integration offers low linewidth solutions including ECLs[8–11], self-injection locking lasers (SIL)[12,13,38], and stimulated Brillouin scattering (SBS) lasers[14–19]. State of the art integrated ECLs are capable of output powers in the range of 20 mW with the fundamental linewidth on the order of 10 Hz[8,9]. Large mode volume table-top and integrated ECLs can approach Hz-level fundamental linewidths and can employ external lab-scale or integrated[26] resonator reference cavities in a reflection mode to further reduced integral linewidths, where the reference is a cold-cavity and the resulting high frequency noise defined by the single mode laser that is stabilized to one reference cavity resonance. With these cold-cavity stabilized lasers, the large reference cavity mode volume is used to reduce the thermorefractive noise (TRN) floor and associated close to carrier and mid-carrier offset frequencies (e.g. 1 Hz to 1 kHz). However, using a cold reference cavity in the transmission mode to reduce fundamental linewidth, e.g. offset frequencies beyond 100 kHz, is extremely difficult since the low linewidth cavity technical noise translates to the high frequency noise.

Lasing based on nonlinear optical feedback, such as integrated SBS and SIL lasers, are important approaches to reducing fundamental linewidth, capable of producing sub-Hz linewidths [14–17]. To date, it has been difficult to achieve both linewidth reduction and output power increase due to laser physics including mode saturation, the onset of higher order modes, and limited cavity mode volume and the resulting modified Schawlow-Townes linewidth (STLW) and TRN limits. In terms of the nonlinear feedback physics, there is a fundamental difference between SIL and SBS. The SIL laser augments an already lasing single mode, e.g. DFB or DBR laser, or multimode Fabry

Perot (FP) laser with a high reflectivity feedback cavity, to selectively reduce the existing spontaneous in the single mode laser for the former, or drive the already lasing multi-mode FP laser into single mode operation through coherent collapse for the later. Notably, SIL utilizes highly nonlinear feedback to provide noise suppression in an already lasing semiconductor resonator. SBS lasers utilize a fundamentally different nonlinear mechanism, where an external laser is used to generate phonons in a resonator and reduce optical frequency noise through highly nonlinear feedback between input and output photons and phonons. Yet, achieving both high output power and narrow linewidths in integrated SBS lasers has been challenging. To date, these lasers employ small cavity mode volumes with an FSR set by the phase matching condition between cavity resonance and the Brillouin frequency shift. The requirement for single mode operation has driven SBS laser cavity designs to support one or multiple FSR per Stokes frequency shift from the pump where the Brillouin gain bandwidth overlaps with only a single cavity resonance[14–17]. This design inherently leads to cascaded emission, which limits the output power and linewidth narrowing[23]. Designs that inhibit the onset of second order Stokes lasing have been employed to decrease the linewidth and increase the single mode output power[20–22]. However, in these designs the first-order Stokes (S1) laser emission grows modestly— scaling with the square-root of the pump power—limiting the output power and the linewidth[22,23]. Additionally, the small mode volume further limits the linewidth through the TRN floor which scales inversely with the optical mode volume[24,25] and ultimately the modified STLW. While increasing the resonator quality factor (Q) of these integrated SBS cavities has enabled linewidth reduction[14,15,17], the resulting low-threshold coupled with cascaded emission limit the linewidth and the output power. Therefore, new solutions are needed to overcome these limits and enable scalable reduction in linewidth and increase in optical output power.

In this work we report a novel integrated SBS laser that can simultaneously achieve high output power and low fundamental linewidth into a spectrally pure single frequency mode. Our laser utilizes an active meter-scale coil waveguide resonator to drive down the linewidth and drive up optical power through an increase in intracavity photons, reduction in the TRN floor, and an increase in the S1 optical power saturation level. The laser outputs a single mode 31 mHz fundamental linewidth with 41 mW optical power and strong single mode operation with 73 dB sidemode suppression ratio (SMSR) using an active 160 million intrinsic Q meter-scale coil-resonator cavity (4-meters). We further demonstrate that this laser can be Vernier tuned across a 22.5 nm range. This work differs fundamentally from prior works, in terms of operation and underlying physics, that employ meter-scale cold-cavity references for integral linewidth reduction[26]. In the present work, the resonator is the nonlinear laser cavity itself, where Brillouin physics inside the cavity provides extreme down selection of both the pump laser photons and the vacuum driven spontaneous emission modes. The Brillouin nonlinear dynamics is so selective that vacuum driven spontaneous modes within the primary resonance and all other cavity resonances that overlap the Brillouin gain are excluded from utilizing pump photons for scattering except for the very select modes. This property is experimentally measured and demonstrated.

Remarkably, although the Si$_3$N$_4$ waveguide meter-scale coil resonator is a multimode cavity with 48.1 MHz FSR that supports 5 longitudinal cavity modes across the Brillouin gain bandwidth, the strong nonlinear Brillouin photon-phonon feedback process demonstrates such high selectivity that the output mode measures greater than 70 dB SMSR. This process is defined by the strength and bandwidth selectivity of the dominant Brillouin grating that scatters the output Stokes light. In other words, the highly nonlinear physics of the dominant Brillouin grating (phonon) formation in the coil resonator itself "steals" all of the pump photons to the drive the single lasing mode. This "winner take all," out-competes the vacuum driven spontaneous modes withing the primary resonance and the other resonances that overlap the Brillouin gain bandwidth. This lasing principle is fundamentally different than approaches that use a meter-scale coil resonator as an external stabilization cavity to reduce the close to carrier noise for a separate laser operating above threshold[26] that does not drive down fundamental linewidth and does not provide increase in optical output power. As a result the Brillouin dynamics in our large mode volume, ultra-high Q cavity, can scale both linewidth and power. We discuss that a path forward with increasing cavity length can be extended to drive the fundamental linewidth down to 1 mHz or lower and output power to above 1 Watt. Owing to its compact footprint and efficiency, this approach to narrow linewidth and high-power integrated lasers can enable portable new forms of precision applications spanning from visible to NIR wavelengths.

## Results

**Principle of operation.** The principle of operation and the large mode volume SBS laser configuration are illustrated in Fig. 1. A 4-meter long laser cavity is implemented as an ultra-low loss silicon nitride (Si$_3$N$_4$) waveguide (Fig. 1**a**) coiled into a bus-coupled resonator on a chip smaller than a square centimeter (Fig. 1**b**). A critical point is that the Brillouin laser cavity can be implemented in a complex waveguide geometry such as the coil resonator owing to the physics of this type of Brillouin laser, where the photon-phonon scattering interaction occurs over the length of the resonator without using phonon guiding[16,23]. The absence of phonon waveguiding leads to a broad Brillouin gain spectrum of approximately 250 MHz at 1550 nm, which spans approximately 4 coil cavity FSRs (Fig. 1**c**). The physics of this Brillouin laser does not require acoustic guiding, allowing the resonator design to focus on the waveguide design for low propagation loss. Below threshold, the pump is spontaneously back-scattered by a thermal population of phase-matched phonons into the four optical resonances that lie within the Brillouin gain bandwidth (illustrated by the orange, red, green, and purple acoustic gratings in Fig. 1**d**). The scattered power in these four optical modes continues to grow with increasing pump until threshold is met for the mode with the largest optical gain (S1). Above threshold the intracavity pump power is clamped (Fig. 1**e**), and only the lasing mode S1 increases with supplied power (purple in middle of Fig. 1**f**). Consequently, the other sub-threshold modes within the SBS gain bandwidth that are driven by the pump are clamped, remaining fixed in power (orange, red, green in center of Fig. 1**f**). The result is illustrated in Fig. 1**d**, 1**e**, 1**f** with single mode operation in a large mode volume and traditionally multimode resonator. The nonlinear feedback process of stimulated Brillouin scattering is so dominant, the lasing mode also takes away pump photons from the build-up of spontaneous modes (acoustic gratings or phonons) within the lasing resonance itself, which leads to very strong linewidth narrowing that leads to orders of magnitude smaller laser (S1) linewidth than the cavity resonance[23].

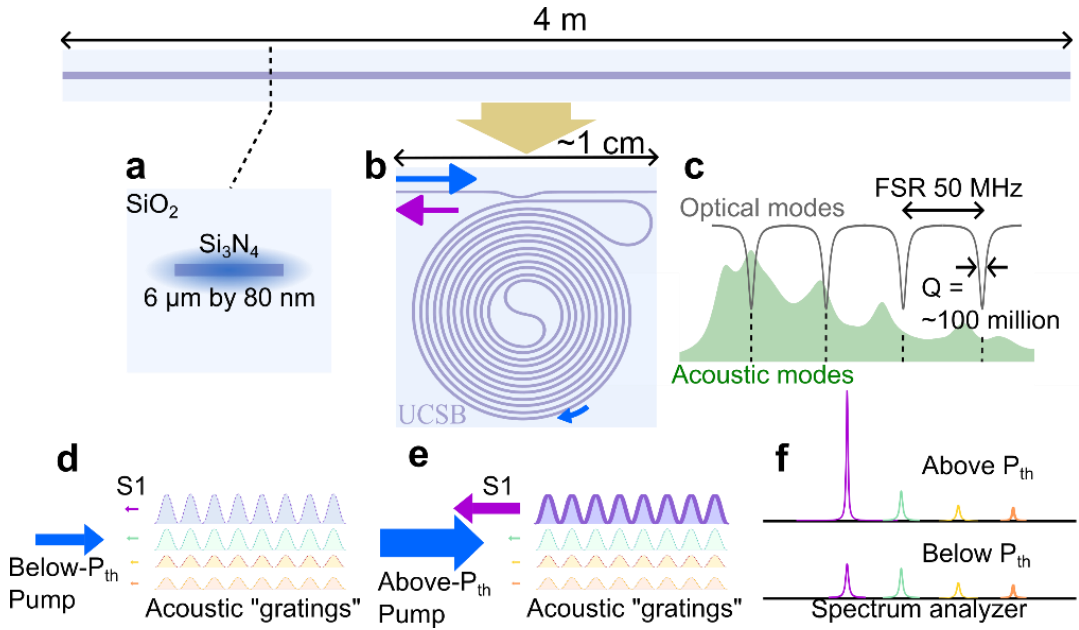

**Fig. 1. Large mode volume SBS laser design and operation principles. a** 6 $\mu$m by 80 nm $Si_3N_4$ waveguide design supports the fundamental TE mode with a minimum bending radius of ~1 mm and propagation loss of ~0.2 dB/m in C and L band. **b** Large mode volume SBS laser implemented in a 4-meter-coil resonator. **c** Coil resonator with 50 MHz FSR overlaps with the broad spontaneous acoustic mode continuum of the Brillouin gain bandwidth (green). **d, e** Principle of single mode operation in a multi-mode laser cavity for below (**d**) and above (**e**) Brillouin lasing threshold. Below threshold the pump scatters uniformly in gratings formed at each of the cavity resonance that overlap with the Brillouin gain. Above threshold, the grating at the resonator mode that overlaps with the Brillouin gain peak rapidly builds up leading to single mode lasing in a large mode volume, normally multi-mode laser resonator. **f** Lasing modes below and above threshold. $SiO_2$, silicon dioxide. $Si_3N_4$, silicon nitride. FSR, free spectral range. Q, quality factor. S1, first-order Stokes. $P_{th}$, optical threshold power.

**Coil SBS laser design.** The 4-meter-coil resonator design sets the foundation of SBS laser. For a large mode volume SBS laser, due to the cascaded Stokes emission nature of SBS lasers in general, it is desirable to operate the pump power just below the second-order Stokes (S2) threshold. For the design demonstrated here, the S2 threshold is expected to be 4 times the S1 threshold. Driving the pump up to just below the onset of S2 lasing is the point where the S1 fundamental linewidth reaches a minimum[15,22,23]. The cavity external coupling Q ($Q_{ex}$), loaded Q ($Q_L$) and the cavity Brillouin gain ($\mu$) determine the S1 threshold, output power, and the minimum fundamental linewidth[15,22,23]. The S1 threshold ($P_{th}$) and the S1 output power at its power clamping point are linearly proportional to the cavity length while the corresponding minimal fundamental linewidth ($\Delta \nu_{min}$) reduces linearly with the cavity length ($L$). In this mode of operation, it is desirable to increase the optical mode volume by increasing the cavity length, which in turn increases the S1 output power and reduces the fundamental linewidth. The waveguide loss and coil resonator Q are important parameters, since SBS threshold is reduced as the cavity Q increases.

For the coil resonator we employ a silica clad 6 $\mu$m wide by 80 nm thick $Si_3N_4$ waveguide design. This waveguide supports the fundamental transverse-electric ($TE_0$) mode with moderate waveguide confinement, propagation loss around 0.2 dB/m in C and L band, and a critical bending radius less than 1 mm (see the waveguide design in Supplementary Fig. S1). By leveraging this tight bending radius, a 4-meter-long waveguide resonator can be realized in a penny-size footprint (see the photograph in Supplementary Fig. S1). The resulting coil resonator has a 48.1 MHz FSR and 160 million intrinsic Q in the C and L band (see the waveguide loss and resonator Q measurements in Supplementary Fig. S3). Within the coil, neighboring waveguides are spaced by 30 $\mu$m and the minimum radius of curvature is ~1.8 mm to avoid bending losses. A comprehensive review of the design considerations of coil resonators can be found in previous work[26,27]. The device fabrication details can be found in Methods as well as in previous work[17,28]. The $TE_0$ mode waveguide losses are measured by spectral scanning the coil resonator with the laser detuning calibrated by a fiber Mach-Zehnder interferometer (MZI). The lowest loss is measured to be ~0.2 dB/m from 1550 nm to 1630 nm in L band while below 1550 nm the loss increases due to the N-H absorption at 1520 nm[27,29]. For better fiber-to-chip coupling, the bus waveguide is tapered from 6 $\mu$m to 11 $\mu$m. Ultra-high numerical aperture (UHNA) fibers with a mode field diameter of 4.1 $\mu$m are used to fiber-pigtail the tapered waveguide facets. The mode-overlapping simulation (shown in Supplementary Fig. S2) shows a theoretical coupling loss between the UHNA fiber and the 11 $\mu$m tapered waveguide of 1.9 dB per facet and 3.8 dB total. The fiber pigtailed connection results in a total coupling loss of 5.8 dB (see photo of packaged coil resonator in Supplementary Fig. S1).

**Coil SBS laser characterization.** The SBS laser characterization is featured with the achievement of high output power and narrow fundamental linewidths. The results shown in Fig. 2 were obtained using a widely tunable external cavity laser as the SBS pump which is amplified by an erbium doped fiber amplifier (EDFA) at the resonator, then Pound-Drever-Hall (PDH) locked to the coil resonator near 1570 nm. The S1 SBS emission propagating in the opposite direction to the pump is dropped by a fiber circulator located at the chip input. The S1 emission is photo-mixed with the pump on a fast photodetector to generate a beatnote to resolve the S1 power spectrum in the radio-frequency domain on an electric spectral analyzer (ESA). Just below threshold, in the spontaneous Brillouin emission regime, multiple Stokes tones are observed, which resembles the simulated Brillouin gain profile (see the method of the Brillouin gain simulation in Ref. [16,18]). Above threshold, we observe rapid emergence of the main Stokes lasing mode (S1) with a 73 dB SMSR (Fig. 2). Above threshold, all increases in pump energy are directed to the primary lasing mode, enabling this large contrast between the power in the lasing mode and the side modes. Also observed is the further Brillouin linewidth reduction that occurs at the primary cavity mode. A coupled-mode model (described in Supplementary Note 2) shows that this single-mode lasing results from clamping of the pump once threshold is met for the Stokes mode with the largest gain, which inhibits competing optical modes from reaching threshold. After confirming the property of single mode lasing, we increase the pump power to increase the S1 output power, which is measured on an optical spectral analyzer (OSA), (see Supplementary Fig. S4). The on-chip S1 output power is calibrated and plotted versus the pump power (inset in Fig. 3 and Supplementary Fig. S4). We measure an S1 threshold of 72 mW and generation of 41 mW on-chip S1 power at a 242 mW pump power, which corresponds to an overall 16.5% efficiency (Supplementary Fig. S4). With the 2.9 dB facet fiber coupling loss, there is ~13 dBm measured optical output power in the fiber.

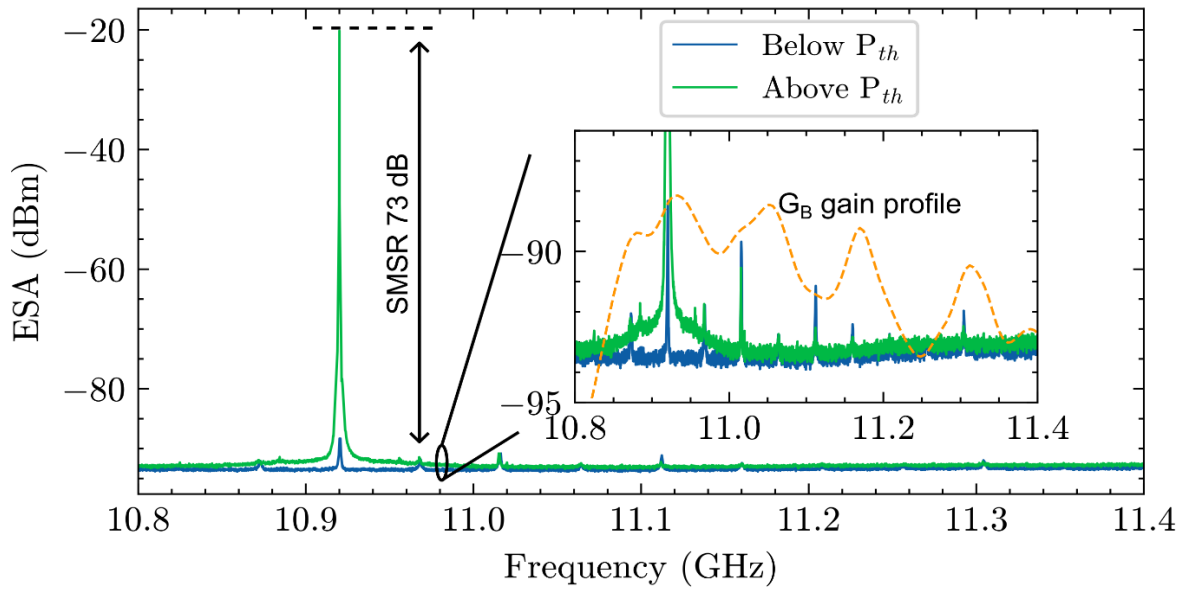

**Fig. 2. Coil SBS laser single mode lasing.** Pump-S1 beatnote spectrum of the SBS laser at 1570 nm on an ESA in spontaneous emission below threshold (blue trace) and stimulated emission above threshold (green trace) demonstrates 73 dB SMSR. Inset shows the Brillouin grating modes at cavity resonances that overlap Brillouin gain spectrum below-threshold (blue trace) and above-threshold (green trace). Measurements show pump energy is driven to the dominant lasing mode above threshold with a reduction in gratings that contribute to side mode power. Also shown is linewidth reduction in the dominant mode above lasing due to successful competition for pump gain within the dominant mode resonator linewidth. ESA, electrical spectrum analyzer. $P_{th}$, SBS threshold. SMSR, side mode suppression ratio, $\boldsymbol{G_B}$, Brillouin gain spectrum.

The laser frequency noise and fundamental linewidth are measured using an unbalanced fiber optic MZI with a 1.037 MHz FSR as an optical frequency discriminator (OFD) and the delayed self-homodyne laser frequency noise measurement method as described in Methods[16,30]. The fundamental linewidth is calculated from the white-frequency-noise floor of the frequency noise spectrum[16,22], $\Delta\nu_F = \pi S_w$. To enable measurement of such low frequency noise and fundamental linewidth we characterize the OFD frequency noise floor that starts to emerge above 10 MHz frequency offset and is primarily from the balanced photodetector used in the OFD (light blue dash in Fig. 3). The red frequency noise curve in Fig. 3 measured 10 mHz$^2$/Hz frequency noise at high frequency offsets and is used to calculate the fundamental linewidth of 31 mHz.

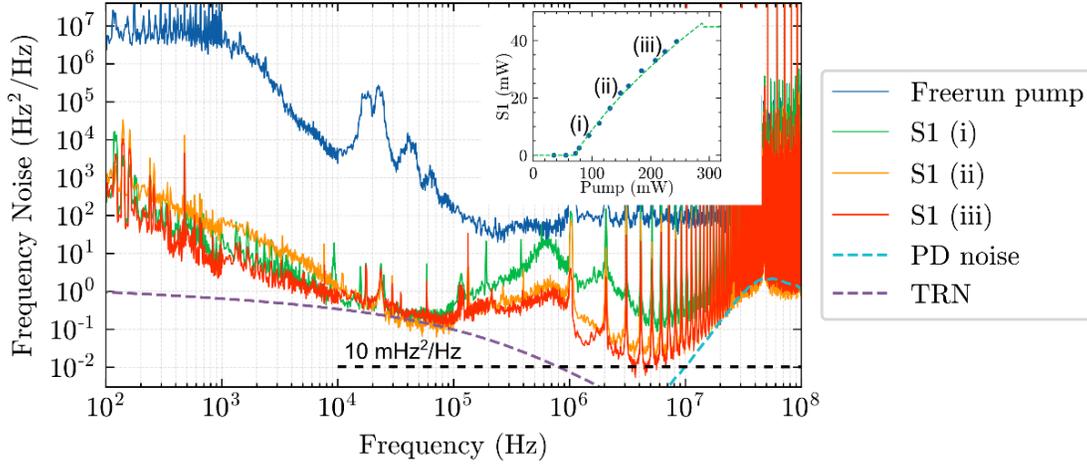

**Fig. 3. SBS laser output power, frequency noise, and fundamental linewidth.** OFD frequency noise measurements for the SBS laser at different pump powers show the minimum fundamental linewidth of 31 mHz, corresponding to a frequency noise of 10 mHz$^2$/Hz. At frequency offset above 10 MHz, the OFD frequency noise measurement is limited by a noise floor from the photodetector. Inset shows the SBS laser on-chip power versus on-chip pump power with a threshold of 72 mW and an output power of 41 mW. Purple dashed curve is the simulated TRN of the coil resonator. At frequency offsets above 10 MHz, the OFD frequency noise measurement is limited by the photodetector noise, as indicated by the light green dashed curve. OFD, optical frequency discriminator. PD, photodetector. TRN, thermo-refractive noise.

It is worth noting that the coil SBS laser frequency noise reaches the coil resonator's TRN limit at frequency offsets from 10 kHz to 100 kHz. Below 10 kHz the S1 frequency noise is dominated by environmental noise such as vibrations and acoustic noise, as well as photothermal noise due to the intracavity power fluctuation induced by the pump laser intensity noise[31,32]. Above 100 kHz and below 1 MHz, there is potentially pump-transferred noise[31], which prevents resolving the fundamental linewidth below 31 mHz. Although we can measure a fundamental linewidth of 31 mHz, the estimated fundamental linewidth in this 4-meter-coil SBS laser operating at the S1 clamping point is lower than the measured value. In previous work[16,33], we measured an SBS laser fundamental linewidth of 0.7 Hz at the S1 clamping point in a 11.83-mm-radius ring resonator with a cavity length of 74.3 mm. Here, the 4-meter-coil cavity length is 54 times the ring resonator cavity length, resulting in fundamental linewidth decrease by 54 times to 13 mHz. The ability to measure low magnitude frequency noise and fundamental linewidth at 0.1 MHz and above is limited by the frequency noise measurement system. Above 10 MHz the OFD frequency noise measurement is limited by the photodetector noise (light blue dashed curve in Fig. 3).

We additionally demonstrate that this approach can be used to realize a tunable SBS laser. The SBS laser can tune every 48 MHz on the coil resonator FSR grid and can further tune continuously by adjusting the coil resonator temperature. Moreover, given that every ~6 nm a pair of optical modes satisfy the SBS phase matching conditions a Vernier tuning scheme can be realized as illustrated in Fig. 4**a**. This behavior is enabled by the small optical FSR and is not possible in conventional bus-coupled ring resonators with FSRs on the order of GHz. This form of tuning is primarily due to the Brillouin shift frequency's dependence on the pump mode wavelength, i.e., $\Omega_B \propto 1/\lambda$ which enables phase matching to achieved for pairs of optical modes separated by $\Omega_B$. This translates to a Vernier effect where the SBS phase matching wavelength points every ~6 nm near 1570 nm, as expressed by $\delta\lambda = \lambda\, \nu_{FSR}/\Omega_B$. The dependence of FSR on wavelength (coil resonator cavity dispersion) is negligible compared to the acoustic frequency dependence on wavelength. The OSA traces of the S1 emission at the on-chip pump power of ~140 mW at different wavelengths from 1550 nm to 1580 nm (Fig. 4**b**) are recorded when S1 reaches a local minimal threshold or local maximal output power. From 1570 nm to shorter wavelengths, the SBS threshold increases due to increased waveguide losses and decreased cavity Qs (see Supplementary Fig. S3 and S5). Although above 1570 nm the SBS threshold is estimated to decrease, the C-band EDFA does not provide optical gain and amplification above 1580 nm to pump the SBS laser. With the technique of two-point coupling that provides resonator coupling in large wavelength range, it is possible in the future work to achieve an SBS laser in one single coil waveguide resonator at multiple wavelength bands such as C, L, O bands, and even visible light[19,27].

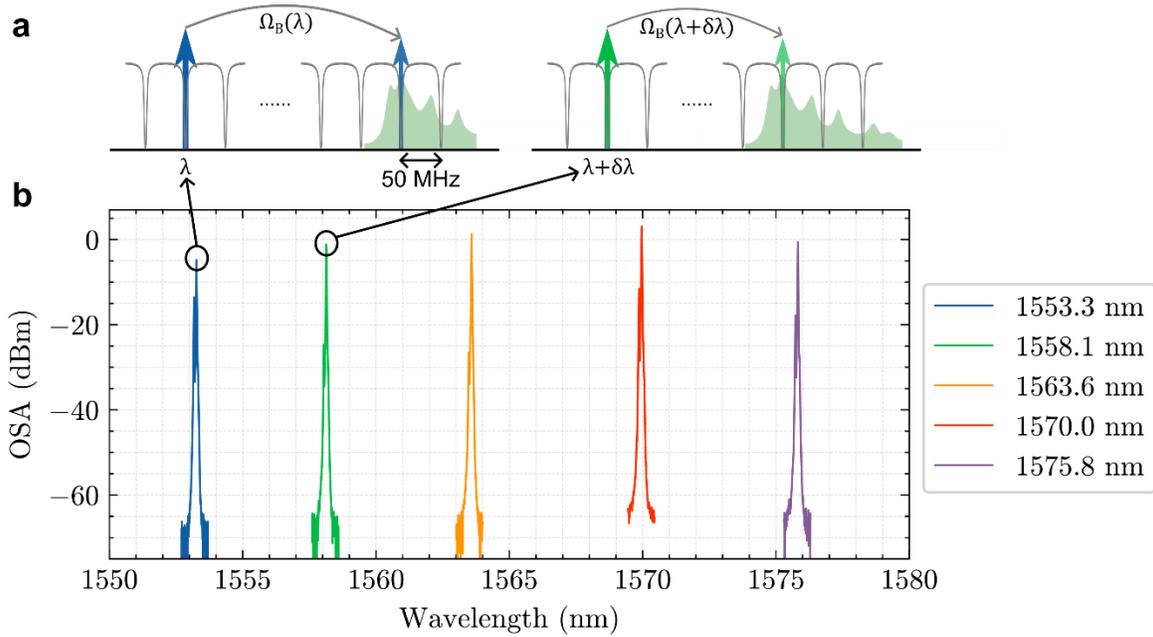

**Fig. 4. Wide-wavelength-range coil SBS laser in C and L band. a**, The SBS phase matching condition is satisfied every ~6 nm, as the Brillouin shift $\Omega_B$ that is linearly proportional to the optical frequency changes by 50 MHz when the pump wavelength changes by ~6 nm around 1570 nm. **b**, S1 emission recorded on the OSA reaches a local minimal threshold every ~6 nm in C and L band. OSA, optical spectrum analyzer. $\Omega_B$, Brillouin shift frequency.

## Discussion and Conclusion

We demonstrate a photonic integrated laser that can simultaneously achieve ultra-narrow linewidth single mode operation and high emitted power. The design employs SBS lasing in a large mode volume, meter-scale coil resonator cavity to achieve 31 mHz fundamental linewidth, 41 mW output power, a 73 dB SMSR, and Vernier tuning across a 22.5 nm range. This performance corresponds with greater than 5 orders magnitude frequency noise reduction from the free running pump laser over a wide bandwidth. These results are made possible with a 4-meter-long coil resonator cavity with 160 million intrinsic Q, increasing in the S1 threshold and emitted power, thereby reducing the fundamental linewidth. The large optical volume also leads to a reduced TRN floor for the coil cavity, minimizing technical forms of thermal noise[31,32].

Remarkably, the Brillouin lasing physics allows the use of a large mode nonlinear photon-phonon resonator with multi-FSR per gain bandwidth to achieve high SMSR single mode lasing. Using a coupled mode model, and verified experimentally, we show that single mode lasing is promoted for the mode with the largest Brillouin gain. The lasing mode that reaches threshold first prevents vacuum driven spontaneous modes within the primary resonance and from other cavity resonances from reaching stimulated emission. To support these conclusions, we measure the frequency noise from below threshold, through threshold, up to saturation of the S1 mode, demonstrating linewidth reduction and increased coherence. The physics here differ markedly from approaches that employ such resonators as cold-cavity references where the lasing response is defined by the external semiconductor laser that interacts with only one cavity resonance and a PDH loop. In this work, the resonator is the nonlinear laser cavity itself, where Brillouin physics form a process that is so selective, the resulting down selection of pump laser photons as well as vacuum driven spontaneous emission modes in the Brillouin nonlinear dynamics, are capable of forming the 30 mHz linewidth, 40 mW lasing output..

In Fig. 5 we provide a comparison of this laser performance to other integrated and non-integrated micro-optic lasers including SIL and Brillouin lasers[12,16,21,22,34–41]. Also shown in Fig. 5 is that by scaling the coil to 20 meters and beyond, in combination with the Brillouin lasers, this design can decrease the fundamental linewidth down to 1 millihertz fundamental linewidth and output power exceeding 1 Watt. Coil lengths of 4 meters to order 20 meters have been demonstrated today, and longer coils out to 200 meters could be possible using multi-layer silicon nitride waveguides[42,43].

We also have demonstrated that through the Vernier effect, we can tune the laser output wavelength in discrete steps over a range of 22.5 nm . This grid of accessible wavelengths is due to the meter-scale coil FSR of 48.1 MHz and the SBS phase matching condition being satisfied every ~5.5 nm. With the technique of two-point coupling that provides resonator coupling in large wavelength range, it is possible in the future work to achieve an SBS laser in one single coil waveguide resonator at multiple wavelength bands such as C, L, O bands, and even visible light[19,27]. These large mode volume SBS coil lasers are fabricated using a 200-mm CMOS foundry compatible process, demonstrating the potential for low cost, wafer-scale fabrication of compact precision high-performance lasers that can be integrated with other components for systems-on-chip solutions for fiber communications wavelength ultra-low phase applications such as mmWave, RF, and fiber sensing. Additionally, the ability to fabricate these coil resonators and SBS lasers in the visible[18,44] shows a path towards ultra-low fundamental linewidth, high power lasers for portable and increased reliability atomic and quantum experiments and metrology applications.

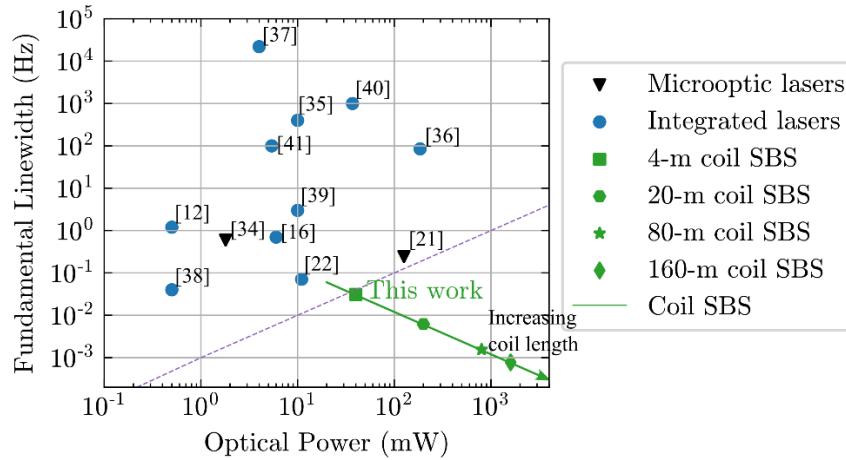

**Fig. 5. Scaling the integrated coil Brillouin laser to 1 mHz fundamental linewidth and greater than 1 Watt output power .** The green line indicates what is possible with the coil SBS lasers, and by increasing coil lengths the SBS laser output power at the S1 clamping point increases linearly with the coil lengths while the fundamental linewidth decreases linearly with the coil length, assuming that the coil resonator Qs stay relatively the same. To reach below 1 mHz fundamental using coil SBS lasers, the coil length needs to be increased by 30 times from 4 meters to 120 meters, and the output power can be increased to 1200 mW.

## Materials and methods

**Fabrication process.** A 15-μm-thick thermal oxide lower cladding layer is grown on a 200-mm diameter silicon substrate wafer. A 75-nm-thick $Si_3N_4$ film is deposited on the thermal oxide using low-pressure chemical vapor deposition (LPCVD), followed by a standard deep ultraviolet (DUV) photoresist spinning, DUV stepper patterning and dry etching in an inductively coupled plasma etcher using $CHF_3/CF_4/O_2$ chemistry. After etching, Following the etch, a standard Radio Corporation of America (RCA) cleaning process is applied. An additional $Si_3N_4$ thin layer is deposited followed by a 30-minute and 1100 °C anneal in an oxygen atmosphere. Lastly, a 6-μm-thick silicon dioxide upper cladding layer is deposited using plasma-enhanced chemical vapor deposition (PECVD) with tetraethoxysilane (TEOS) as a precursor, followed by a final two-step anneal at 1050 °C for 7 hours and 1150 °C for 2 hours.

**Laser frequency noise and fundamental linewidth measurements.** A 200-m fiber MZI with a FSR of 1.03 MHz is used as an optical frequency discriminator (OFD)[16,22,30] with the two fiber outputs detected with a Thorlabs balanced photodetector (PDB 450C, transimpedance gain $10^4$ V/A) and the balance detection signal sampled by a Keysight DOXS1024 digital oscilloscope at different sampling rates ($10^4$ Sa/s, $10^5$ Sa/s, $10^6$ Sa/s, $10^7$ Sa/s, $10^8$ Sa/s) with 10,000 sampling points at each sample rate. The power spectrum of the balanced detection signal $S_{BPD}(f)$ is converted into a laser noise spectrum $S_\nu(f)$ by[16,22],

$$S_\nu(f) = S_{BPD}(f)\left(\frac{f}{\sin(\pi f \tau_D)\,V_{pp}}\right)^2,$$

where $1/\tau_D$ is the fiber MZI FSR and $V_{pp}$ is the peak-to-peak voltage of the fiber MZI response over a full FSR detuning frequency. The balanced detection needs to be as close to perfect balancing as possible to minimize potential laser intensity noise to frequency noise conversion, and the balance-detected power is maximized for a reduced laser noise measurement noise floor, as described in more details in Ref. [22].

## Acknowledgements
This material is based upon work supported by DARPA GRYPHON, under Award Number HR0011-22-2-0008, and ARO AMP under Award Number W911NF-23-1-0179. The views and conclusions contained in this document are those of the authors and should not be interpreted as representing official policies of DARPA or the U.S. Government. We thank Jim Nohava, Joe Sexton, Jim Hunter, Dane Larson, Michael DeRubeis, and Jill Lindgren at Honeywell for their contributions to mask design and sample fabrication. We also thank Kwangwoong Kim at Nokia Bell Labs for fiber-pigtailing and packaging the coil resonator.

## Author contributions
K. L., R. O. B. and D. J. B. prepared the manuscript. K. L. and K. D. N. designed and fabricated the coil resonator. K. L. designed and implemented the packaging. R. O. B. simulated the Brillouin gain response and performed the theoretical modeling and analysis of Brillouin lasing. K. L. performed the SBL lasing demonstration with the OSA, ESA, and frequency noise measurements. D.J.B. supervised and led the scientific collaboration.

## Data availability
The data and code that support the plots and other findings of this study are available from the corresponding authors upon reasonable request.

## Conflict of interest
Dr. Blumenthal's work has other work funded by Infleqtion. Dr. Blumenthal has consulted for Infleqtion and received compensation, is a member of the scientific advisory council, and owns stock in the company. K. Liu, K. Nelson, and R. O. Behunin declare no potential conflict of interest.

## Supplementary information
Supplementary Information is available.

# References

1. Kikuchi, K. Fundamentals of Coherent Optical Fiber Communications. *J. Light. Technol.* **34**, 157–179 (2016).

2. Mecozzi, A. *et al.* Use of Optical Coherent Detection for Environmental Sensing. *J. Light. Technol.* **41**, 3350–3357 (2023).

3. Lihachev, G. *et al.* Low-noise frequency-agile photonic integrated lasers for coherent ranging. *Nat. Commun.* **13**, 3522 (2022).

4. Xu, C. *et al.* Sensing and tracking enhanced by quantum squeezing. *Photonics Res.* **7**, A14 (2019).

5. Ludlow, A. D., Boyd, M. M., Ye, J., Peik, E. & Schmidt, P. O. Optical atomic clocks. *Rev. Mod. Phys.* **87**, 637–701 (2015).

6. Sun, S. *et al.* Integrated optical frequency division for microwave and mmWave generation. *Nature* (2024) doi:10.1038/s41586-024-07057-0.

7. Kudelin, I. *et al.* Photonic chip-based low-noise microwave oscillator. *Nature* (2024) doi:10.1038/s41586-024-07058-z.

8. Fan, Y. *et al.* Hybrid integrated InP-$Si_3N_4$ diode laser with a 40-Hz intrinsic linewidth. *Opt. Express* **28**, 21713 (2020).

9. Wu, Y. *et al.* Hybrid integrated tunable external cavity laser with sub-10 Hz intrinsic linewidth. *APL Photonics* **9**, 021302 (2024).

10. Morton, P. A. & Morton, M. J. High-Power, Ultra-Low Noise Hybrid Lasers for Microwave Photonics and Optical Sensing. *J. Light. Technol.* **36**, 5048–5057 (2018).

11. Grillanda, S. *et al.* Hybrid-Integrated Comb Source With 16 Wavelengths. *J. Light. Technol.* **40**, 7129–7135 (2022).

# Supplementary Information: Large Mode Volume Integrated Brillouin Lasers for Scalable Ultra-Low Linewidth and High Power

Kaikai Liu[1], Karl D. Nelson[2], Ryan O. Behunin[3, 4], Daniel J. Blumenthal[1*]

[1]Department of Electrical and Computer Engineering, University of California Santa Barbara, Santa Barbara, CA, USA
[2]Honeywell Aerospace, Plymouth, MN, USA
[3]Department of Applied Physics and Materials Science, Northern Arizona University, Flagstaff, Arizona, USA
[4]Center for Materials Interfaces in Research and Applications (¡MIRA!), Northern Arizona University, Flagstaff, AZ, USA
[*]Corresponding author (danb@ucsb.edu)

# Supplementary Note 1: 4-meter-coil resonator design and testing

Table S1 summarizes the main characteristics of the stimulated Brillouin scattering (SBS) lasers such as output power, efficiency and fundamental linewidth. Normally due to cascaded emission, the SBS laser operates with the pump power of $P_{in} = 4P_{th}$ to achieve the minimum fundamental linewidth $\Delta \nu_{ST} = \mu(n_{th} + 1)/2\pi$, where $\mu$ is the cavity Brillouin gain rate per photon and $n_{th}$ is the thermal occupation number of the acoustic mode, defined in Ref. [1]. With S2 suppression, the S1 output power scales with the square-root of pump power and thus the efficiency drops at high pump power. Whereas, with S3 suppression, the S1 output power scales linearly with the pump power.

| Table S1 SBS laser in (1) cascaded, (2) S2-suppressed, (3) S3-suppressed emissions. | | | |
|---|---|---|---|
| | **Output power** | **Efficiency** | **Linewidth (*)** |
| (1) Cascaded emission at $P_{in} = 4P_{th}$ | $P_{S1} = \frac{Q_L^2}{Q_{ex}^2} 4P_{th}$ | $\eta_{S1} = \frac{Q_L^2}{Q_{ex}^2}$ | $\Delta \nu_{ST} = \frac{\mu(n_{th}+1)}{2\pi} \sim \frac{1}{L}$ |
| (2) S2-suppressed emission | $P_{S1} = 4\frac{Q_L^2}{Q_{ex}^2} \left( \sqrt{P_{th} P_{in}} - P_{th} \right)$ | $\eta_{S1} \cong 4\frac{Q_L^2}{Q_{ex}^2} \sqrt{P_{th}/P_{in}}$ | $\Delta \nu_{ST} = \frac{\hbar \omega^3 (n_{th}+1)}{4\pi Q_L Q_{ex} P_{S1}}$ |
| (3) S3-suppressed emission | $P_{S2} = \frac{Q_L^2}{Q_{ex}^2}(P_{in} - 4P_{th})$ | $\eta_{S2} \cong \frac{Q_L^2}{Q_{ex}^2}$ | $\Delta \nu_{ST} = \frac{\hbar \omega^3 (n_{th}+1)}{4\pi Q_L Q_{ex} P_{S2}}$ |

**Table S1. Output power, conversion efficiency, and fundamental linewidth for (1) cascaded, (2) S2-suppressed, (3) S3-suppressed SBS lasers.** $P_{in}$ is the input pump power, $P_{th}$ is the S1 threshold, $Q_{ex}$ and $Q_L$ are the cavity external coupling and loaded Qs, and $n_{th}$ is the thermal occupation number of the acoustic mode. (*The optical mode thermal occupation number $N_{th}$ is negligible at room temperature and neglected in the term of $(n_{th} + 1)$, $(n_{th} + N_{th} + 1) \cong (n_{th} + 1)$. If the SBS linewidth enhancement factor $\alpha$ is non-zero[2], the linewidth expressions can be multiplied by $(1 + \alpha^2)$. The derivation of these equations can be found in previous work[1,3,4].

The fundamental TE mode in the 6 $\mu$m by 80 nm waveguide shown in Fig. S1**a** can have a bending radius down to 1.0 mm without inducing significant bending loss, shown in Fig. S1**b**. The small bending radius of this waveguide design enables packing the 4-meter-long coil waveguides in a small area of 1.2 cm by 1.2 cm, as shown in Fig. S1**c**. The 4-meter-coil resonator device is fiber pig-tailed using UHNA fibers for better matching with the waveguide mode, and packaged in a metal enclosure, shown by a photo of the device in Fig. S1**c**. The mode matching simulation on the fiber modes and the tapered waveguide mode at different tapering widths is shown in Fig. S2. The resonator bus waveguide uses a waveguide taper of 11 $\mu m$ in this current design, and future designs will use 1.0 or 1.5 $\mu$m tapering width for better coupling.

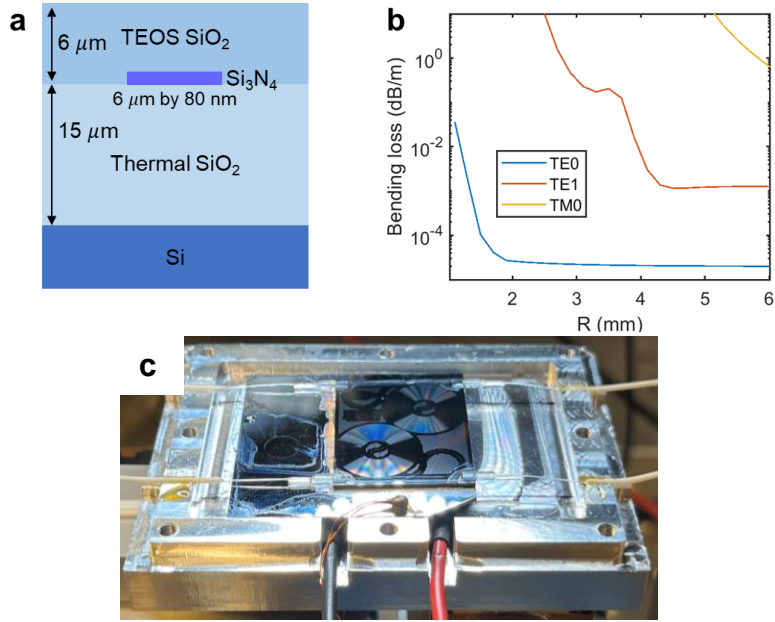

**Fig. S1. Waveguide design and bending loss simulation. a** The 6 $\mu$m by 80 nm thick $Si_3N_4$ waveguide with 15-$\mu$m-thick thermal oxide lower cladding and 6-$\mu$m-thick TEOS-PECVD deposited oxide upper cladding supports the $TE_0$, $TE_1$ and $TM_0$ modes at C and L bands. **b** The minimum bending radius from Lumerical MODE simulations can be as small as 1.0 mm without incurring significant bending loss in the $TE_0$ mode. **c** A photograph shows a coil resonator in a metal package with UHNA fibers attached to the bus waveguides and a thermal electric cooler in between the chip and the metal package for temperature regulation.

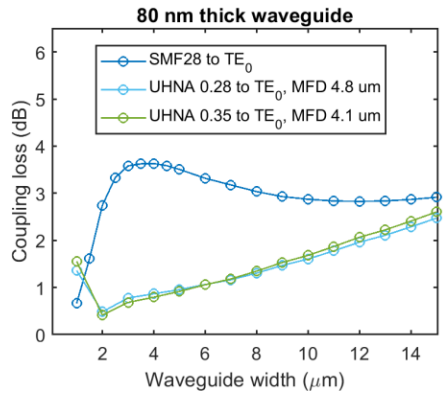

**Fig. S2. Mode-overlapping and coupling efficiency simulation for the input and output waveguide with different widths and different types of fibers.** The waveguide mode is simulated in Lumerical MODE at different waveguide widths and the mode-overlapping is calculated in MODE between the waveguide mode and fiber mode in different types of fibers. Note at the time of the design, the 11 $\mu$m waveguide taper width was chosen for optimizing the coupling efficiency for single-mode fiber 28 rather than the UHNA fibers and a smaller taper width such as 1.5 $\mu$m or 1.0 should be adopted in future taper designs.

The coil resonator Q and linewidth are characterized by spectral scanning the resonator using a widely tunable laser as a probe laser along with the laser detuning calibration from a unbalanced fiber MZI with an FSR of 5.87 MHz. At 1600 nm, the FSR is measured to be 48.1 MHz and the intrinsic and loaded linewidths are 1.1 MHz and 2.2 MHz, respectively, corresponding to 0.15 dB/m propagation loss (Fig. S3**a**). Such resonator Q and linewidth measurements are performed using multiple widely tunable lasers to cover the wavelength range from 1260 nm to 1650 nm (Fig. S3**b**, S3**c**).

The SBS laser output from the circulator is measured on an optical spectral analyzer to ensure there is only S1 emission, increasing with the increasing pump power (Fig. S4**a**). With the calibration of fiber to chip coupling loss, and fiber connection losses, the measured S1 output power from OSA combining with the recorded pump power yield the results of the on-chip S1 power versus the on-chip pump power (Fig. S4**b**), showing the S1 threshold of 72 mW.

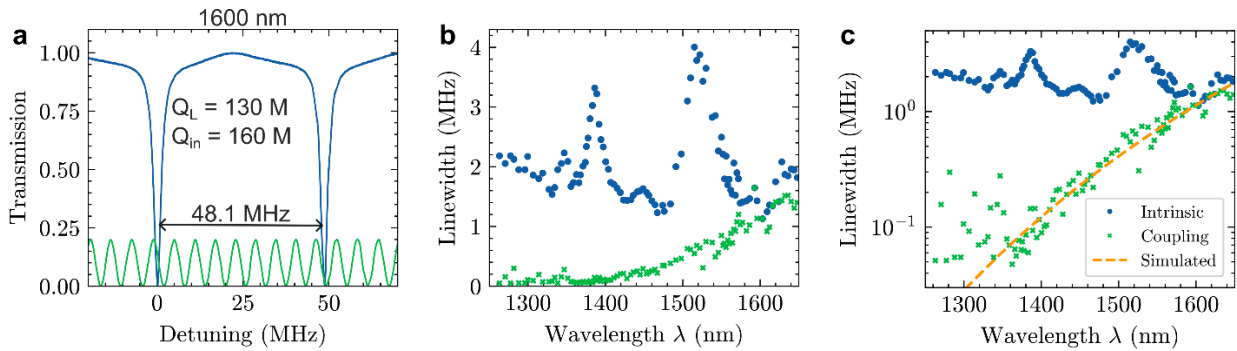

**Fig. S3. Waveguide propagation loss spectrum**. **a** By spectrally scanning the coil resonator using several Santec tunable lasers along with a 5.87 MHz FSR fiber unbalanced MZI for optical frequency detuning calibration, the coil resonator intrinsic loss rate $\gamma_{in}$ and bus-resonator coupling loss rate $\gamma_{ex}$ are measured. At 1600 nm, the waveguide propagation loss is measured to be 0.16 dB/m, the intrinsic Q is 160 million, and the FSR is measured to be 48.1 MHz. **b, c** The resonator intrinsic and coupling linewidths are measured from 1260 nm to 1650 nm and plotted in linear scale (**b**) and in log scale (**c**) and used to estimate the coil waveguide propagation loss by $\alpha = \gamma_{in} n_g / c$[5,6].

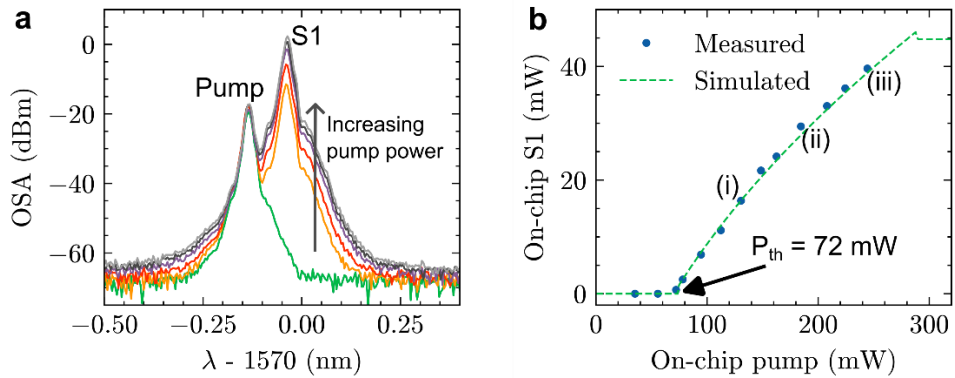

**Fig. S4. Coil SBS threshold and output power. a** The SBS laser S1 and pump power measured on an OSA with increasing pump power. **b** On-chip S1 power versus the pump power measures a 72 mW threshold.

In the main results, the S1 emission is recorded at 1553.3 nm, 1558.1 nm, 1563.6 nm, 1570.0 nm, 1575.8 nm, which correspond to the wavelength points where the Brillouin phase matching condition is satisfied and the S1 threshold is with the reach of the EDFA maximum output power. With the measured resonator linewidths at different wavelengths, it is possible to predict the S1 threshold at all these phase matching wavelength points using the equations in Table S1. Figure S5**a** shows the interpolation of the resonator intrinsic and coupling loss rates, which is used to calculate the S1 threshold versus wavelength (Fig. S5**b**). The increased resonator loss around 1520 nm incurs a significant increase in S1 threshold from 1500 nm to 1550 nm, and low threshold SBS is possible from 1550 nm to 1600 nm as well as near 1470 nm, theoretically.

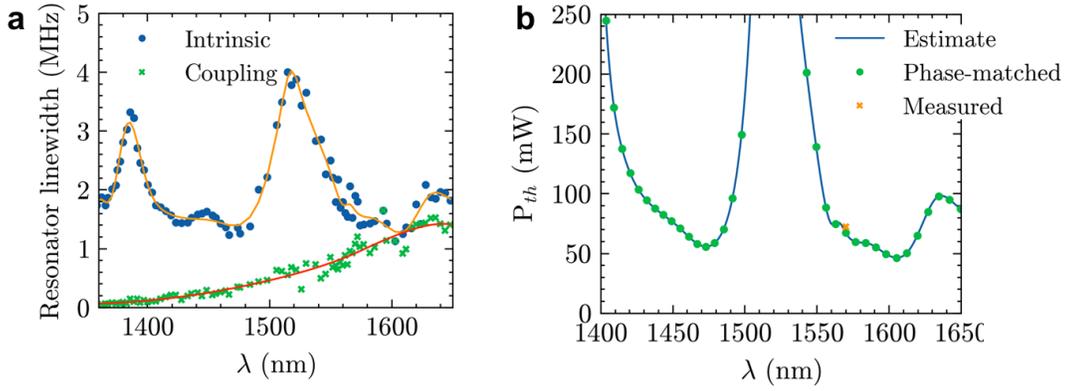

**Fig. S5. Coil SBS threshold estimation at different wavelengths. a** The coil resonator intrinsic and coupling loss rates are measured by a widely tunable laser and interpolated with smoothen curves. **b** Coil cavity Brillouin gain is calculated to be $\mu = 0.174$ mHz using the measured 72 mW SBS threshold at 1570 nm. Assuming the same value for $\mu$ and cavity FSR ($\nu_{FSR} = 48.1$ MHz) at different wavelengths, using the interpolated resonator loss rates, the blue curve shows the continuous estimation of SBS threshold, and the green dots indicates the Brillouin-shift phase matched wavelengths points $\{\lambda_n : n \times \nu_{FSR} = \Omega_{ac,0} \frac{1570}{\lambda_n}\}$.

## Supplementary Note 2: SBS laser coupled mode model and single-mode lasing

To interpret the physics underlying the single-mode lasing displayed in Fig. 2 in the main text, we develop and analyze a coupled-mode model of SBS dynamics. This model is a generalization of prior mean-field treatments of SBS laser dynamics[1], accounting for the presence of optical modes with the potential to lase within the Brillouin gain bandwidth. Inspired by the multi-Lorentzian nature of the gain bandwidth (see Fig. 2), we assume that the SBS dynamics is mediated by a collection of phonon modes. Under the condition where the resonator free-spectral range (48 MHz) is greater than the linewidth associated with the mechanical decay $\Gamma/(2\pi)$ (~30 MHz), we assume that each Stokes optical mode couples to a single phonon. With this assumption, the Heisenberg-Langevin equations for this model are given by,

$$\dot{a}_p = -(i\omega_p + \gamma/2)a_p - i\sum_j g^*_j a_j b_j + f_{ext},$$

$$\dot{a}_j = -(i\omega_j + \gamma/2)a_j - ig_j a_p b^\dagger_j,$$

$$\dot{b}_j = -(i\Omega_j + \Gamma/2)b_j - ig_j a_p a^\dagger_j.$$

Analysis of the steady-state solutions reveals how single mode lasing can come about. In steady state, we assume that the amplitudes oscillate with constant magnitude. First, we solve for the complex phonon amplitude,

$$b_j = -\frac{ig_j a_p a^\dagger_j}{i\Delta\Omega_j + \Gamma/2},$$

where $\Delta\Omega_j$ is the difference between the phonon resonant frequency and the frequency of the driving beatnote. Inserting the steady-state solution for $b_j$ into the steady-state equation for $a_j$ we find,

$$0 = -(i\Delta\omega_j + \gamma/2 - \mu_j a^\dagger_p a_p)a_j,$$

where

$$\mu_j = \frac{|g_j|^2}{-i\Delta\Omega_j + \Gamma/2},$$

and the j$^{th}$ mode oscillates at a frequency spaced from resonance by $\Delta\omega_j$. This equation can be solved two ways: below threshold $|a_j| = 0$, and above threshold $|a_j| \neq 0$. Above threshold the steady-state solution requires $(i\Delta\omega_j + \gamma/2 - \mu_j a^\dagger_p a_p) = 0$, giving the clamping condition for the pump,

$$a^\dagger_p a_p = \frac{\gamma}{2\text{Re}[\mu_j]},$$

and the formula for the frequency offset $\Delta\omega_j = \text{Im}[\mu_j] a_p^\dagger a_p = \gamma\, \text{Im}[\mu_j]/(2\text{Re}[\mu_j]) = \gamma \Delta\Omega_j / \Gamma$. The steady-state equations show that the mode with the largest Brillouin gain ($\text{Re}[\mu_k] > \text{Re}[\mu_{j\neq k}]$) will reach threshold first (i.e., when $a_p^\dagger a_p = \gamma/2\text{Re}[\mu_k]$). At threshold the pump clamps and the only viable solution for the remaining Stokes modes is $|a_{j\neq k}| = 0$ because $\left(i\Delta\omega_{j\neq k} + \gamma/2 - \mu_{j\neq k} a_p^\dagger a_p\right) \neq 0$.